**Large-Scale Analysis of the Accuracy of the Journal Classification Systems**

**of Web of Science and Scopus**

Qi Wang

(Department of Industrial Economics and Management, KTH-Royal Institute of Technology, Lindstedtsvägen 30, 11428 Stockholm, Sweden) qi.wang@indek.kth.se

Ludo Waltman

(Centre for Science and Technology Studies, Leiden University, PO Box 905, 2300 AX, Leiden, Netherlands) waltmanlr@cwts.leidenuniv.nl

## Abstract

Journal classification systems play an important role in bibliometric analyses. The two most important bibliographic databases, Web of Science and Scopus, each provide a journal classification system. However, no study has systematically investigated the accuracy of these classification systems. To examine and compare the accuracy of journal classification systems, we define two criteria on the basis of direct citation relations between journals and categories. We use Criterion I to select journals that have weak connections with their assigned categories, and we use Criterion II to identify journals that are not assigned to categories with which they have strong connections. If a journal satisfies either of the two criteria, we conclude that its assignment to categories may be questionable. Accordingly, we identify all journals with questionable classifications in Web of Science and Scopus. Furthermore, we perform a more in-depth analysis for the field of Library and Information Science to assess whether our proposed criteria are appropriate and whether they yield meaningful results. It turns out that according to our citation-based criteria Web of Science performs significantly better than Scopus in terms of the accuracy of its journal classification system.

## 1. Introduction

Classifying journals into research areas is an essential subject for bibliometric studies. A classification system can assist with various problems; for instance, it can be used to demarcate research areas (e.g., Glänzel & Schubert, 2003; Waltman & Van Eck, 2012), to evaluate and compare the impact of research across scientific fields (e.g., Leydesdorff & Bornmann, in press; Van Eck et al., 2013), and to study the interdisciplinarity of research (e.g., Porter & Rafols, 2009; Porter et al., 2008). The two most important multidisciplinary bibliographic databases, Web of Science (WoS) and Scopus, both provide a journal classification system. Previous studies have compared the two databases from various perspectives (for a review of the literature, see Waltman, 2015, Section 3), but a systematic comparison of the accuracy of the journal classification systems of the two databases has not been performed. Thus, this study is focused on examining and comparing the accuracy of the WoS and Scopus journal classification systems.

This paper is organized as follows. We first provide some background information on various classification systems in Section 2. Then, Section 3 defines the criteria we use to identify journals for which classifications may be questionable. Next, Section 4 introduces the data we use and provides some basic statistics on the data. Section 5 reports the results of our analysis. Discussion and conclusions follow in Section 6.

## 2. Background

Many different classification systems of scientific literature are available, both at the level of journals and at the level of individual publications. The following subsections first introduce some currently available mono- and multidisciplinary classification systems and then provide an in-depth discussion on the WoS and Scopus journal classification systems.

### 2.1. Mono-disciplinary classification systems

A mono-disciplinary classification system covers publications in one particular research area and usually provides a classification at a relatively high level of detail. For instance, EconLit, the American Economic Association's electronic bibliography database, offers the Journal of Economic Literature (JEL) classification system. This system provides a classification of publications in the area of economics. Another example can be found in the Chemical Abstracts database, which indexes literature in chemistry and related areas. Chemical Abstracts Service (2015) indicates that it classifies publications into 80 different sections, which can be further aggregated into five broad headings (see also Neuhaus & Daniel, 2008).

Additionally, in the area of medicine, Medical Subject Headings (MeSH) is used by the U.S. National Library of Medicine for indexing and cataloging medical publications (U.S. Nation Library of Medicine, 2015). MeSH categories are organized in a hierarchical structure. The categories are assigned at the level of individual publications (see also Bornmann et al., 2008).

### 2.2. Multidisciplinary classification systems

Compared with mono-disciplinary classification systems, multidisciplinary systems have a broad coverage of research areas. Well-known examples are the WoS and Scopus classification systems, which are further discussed in Subsection 2.3. Unlike mono-disciplinary classification systems, multidisciplinary classification systems typically work at the level of journals rather than individual publications.

Besides the WoS and Scopus classification systems, there are various other multidisciplinary classification systems, for instance the system of Science-Metrix, the system of the National Science Foundation (NSF) in the US, the UCSD classification system, and the system of the Australian and New Zealand Standard Research Classification (ANZSRC). Science-Metrix assigns "individual journals to single, mutually exclusive categories via a hybrid approach combining algorithmic methods and expert judgment" (Archambault et al., 2011, p. 66). The Science-Metrix system includes 176 categories. The NSF system also offers a mutually exclusive classification of journals, but it is more aggregated, consisting of only 125 categories (Boyack & Klavans, 2014). The system is used in the Science & Engineering Indicators of the NSF. A more detailed classification system is the so-called University of California, San Diego (UCSD) classification system. This system, which includes more than 500 categories, has been constructed in a largely algorithmic way. The construction of the UCSD classification system is discussed by Börner et al. (2012). The ANZSRC's Field of Research (FoR) classification system has a three-level hierarchical structure. Journals are classified at the top level and at the intermediate level. Journals can have multiple classifications.

Furthermore, Glänzel and Schubert (2003) designed a two-level hierarchical classification system, which can be applied at the levels of both journals and publications. They adopted a top-bottom strategy; specifically, they first defined categories on the basis of the experience of bibliometric studies and external experts. They then assigned journals and individual publications to the categories. This classification system has for instance been used for measuring interdisciplinarity. In their analysis of interdisciplinarity, Wang et al. (2015) explain that instead of the WoS subject categories they use the more aggregated classification system developed by Glänzel and Schubert (2003).

Algorithmic strategies have been regularly used to construct multidisciplinary classification systems. Algorithmic approaches to construct classification systems at the level of journals have been studied by for instance Bassecoulard and Zitt (1999), Chen (2008), and Rafols and Leydesdorff (2009). A more recent development is the algorithmic construction of classification systems at the level of individual publications rather than journals. Waltman and Van Eck (2012) developed a methodology for algorithmically constructing classification systems at the level of individual publications on the basis of citation relations between publications. Their approach has for instance been used in the calculation of field-normalized citation impact indicators (Ruiz-Castillo & Waltman, 2015).

*2.3. WoS and Scopus classification systems*

WoS, produced by Thomson Reuters, and Scopus, produced by Elsevier, are the two most important multidisciplinary bibliographic databases. They both include various types of sources, such as journals, conference proceedings, and books. Moreover, they both provide a classification system at the level of journals, and they both allow journals to have multiple classifications. However, although WoS and Scopus have many common characteristics, they also differ in various aspects, for instance in their coverage of journals, in their collection policy, and importantly, in their classification of journals. Many studies have compared the two databases. According to a recent literature review (Waltman, 2015, Section 3), previous studies comparing WoS and Scopus are mainly focused on two aspects. One is the coverage of the databases (e.g., López-Illescas et al., 2008; Meho & Rogers, 2008; Mongeon & Paul-Hus, 2016; Norris & Oppenheim, 2007) and the other is the accuracy of the databases when used to assess research output and impact at different levels, ranging from individual researchers to departments, institutes, and countries (e.g., Archambault et al., 2009; Bar-Ilan et al., 2007; Meho & Rogers, 2008; Meho & Sugimoto, 2009). However, no study has systematically compared WoS and Scopus in terms of the accuracy of their journal classification systems.

There is no documentation describing at a reasonable level of detail the methodology used to construct the WoS and Scopus journal classification systems. In the case of WoS, Pudovkin and Garfield (2002) have offered a brief description of the way in which categories are constructed. According to Pudovkin and Garfield, when WoS was established, a heuristic and manual method was adopted to assign journals to categories, and after this, the so-called Hayne-Coulson algorithm was used to assign new journals. This algorithm is based on a combination of cited and citing data, but it has never been published. Besides this, Katz and Hicks (1995), Leydesdorff (2007), and Leydesdorff and Rafols (2009) have indicated that the WoS classification system is based on a comprehensive consideration of citation patterns, titles of journals, and expert opinion. In the case of Scopus, there seems to be no information at all on the construction of its classification system.

It should be mentioned that in the most recent versions of WoS two classification systems are available, namely a system of categories and a system of research areas. The system of categories is more detailed. This system, which is the traditional classification system of WoS and the system on which we focus our attention in this paper, consists of around 250 categories and covers the sciences, social sciences, and arts and humanities. The system of research areas, which has become available in WoS more recently, is less detailed and comprises around 150 areas. Besides these two systems, Thomson Reuters also has a classification system for its Essential Science Indicators. This system consists of 22 subject areas in the sciences and social sciences. It does not cover the arts and humanities.

The Scopus journal classification system is called the All Science Journal Classification (ASJC). It consists of two levels. The bottom level has 304 categories, which is somewhat more than the about 250 categories in the WoS classification system. The top level includes 27 categories.

The WoS and Scopus journal classification systems are frequently used in bibliometric studies, especially the WoS system. However, knowledge about the accuracy of the WoS and Scopus classification systems is very limited. Pudovkin and Garfield (2002, p. 1113) acknowledged that in the WoS classification system "journals are assigned to categories by subjective, heuristic methods. In many fields these categories are sufficient but in many areas of research these 'classifications' are crude and do not permit the user to quickly learn which journals are most closely related." Similarly, Garfield (2006, p. 92) stated that "the heuristic methods used by Thomson Scientific … for categorizing journals are by no means perfect, even though citation analysis informs their decisions." The accuracy of a classification system can seriously influence bibliometric studies. For instance, Leydesdorff and Bornmann (in press) investigated the use of the WoS categories for calculating field-normalized citation impact indicators. They focused specifically on two research areas, namely Library and Information Science and Science and Technology Studies. Their conclusion is that "normalizations using (the WoS) categories might seriously harm the quality of the evaluation". A similar conclusion was reached by Van Eck et al. (2013) in a study of the use of the WoS categories for calculating field-normalized citation impact indicators in medical research areas.

The accuracy of the WoS and Scopus journal classification systems has been a continuous issue of concern, and researchers have therefore explored a number of approaches to improve these classification systems. Glänzel and colleagues have studied several approaches to validate and improve WoS-based classification systems (Janssens et al., 2009; Thijs et al, 2015; Zhang et al., 2010). They have also proposed an improved way of handling publications in multidisciplinary journals (Glänzel et al., 1999a, 1999b). Related to this, López-Illescas et al. (2009) have studied an approach to improve the field delineation provided by categories in the WoS classification system. The SCImago research group has made a number of attempts to improve the Scopus classification system (Gómez-Núñez et al., 2011, 2014, 2016).

There are still no systematic, large-scale analyses of the accuracy of the WoS and Scopus journal classification systems. Given the importance of these classification systems both in bibliometric research and in applied bibliometric work, a comparative study of the accuracy of the WoS and Scopus classification systems is necessary and urgent. Such a study is presented in this paper.

## 3. Methodology

Two types of approaches can be distinguished for assessing the accuracy of journal classification systems. One is the expert-based approach and the other is the bibliometric approach. Applying the expert-based approach at a large scale is challenging. No expert has sufficient knowledge to assess the classification of journals in all scientific disciplines, so a large number of experts would need to be involved. In the case of the bibliometric approach, a further distinction can be made between text-based and citation-based approaches. Text-based approaches could for instance assess whether the textual similarity of publications in journals assigned to the same category is higher than the textual similarity of publications in journals assigned to different categories. However, in this paper, we do not explore this possibility further. Instead, we take a citation-based approach to assess the accuracy of journal classification systems.

Various types of citation relations, such as direct citation relations, bibliographic coupling relations, and co-citation relations, can be used to measure the relatedness of journals. In this paper, we use direct citation relations. This is because "a co-citation or bibliographic coupling relation requires two direct citation relations" (Waltman & Van Eck, 2012, p. 2380), which means that bibliographic coupling and co-citation relations are more indirect signals of the relatedness of journals than direct citation relations. The use of direct citation relations between journals has a long history, going back to work by Narin and colleagues in the 1970s (Carpenter & Narin, 1973; Narin et al., 1972). The use of direct citation relations is also supported by Klavans and Boyack (2015), who study the algorithmic construction of classification systems at the level of individual publications. They conclude that the use of direct citation relations yields more accurate results than the use of bibliographic coupling or co-citation relations.

Of course, we acknowledge that citation relations provide only a partial perspective on the relatedness of journals. As already mentioned, the relatedness of journals can also be assessed using non-citation-based approaches, in particular expert-based approaches and text-based bibliometric approaches. These approaches may provide a different perspective on the relatedness of journals. A purely citation-based approach therefore does not allow us to draw final conclusions on the correctness of the classification of a journal, but it may provide strong signals that certain journals are likely to be misclassified.

Intuitively, our approach based on direct citation relations can be explained as follows. On the one hand, we expect journals in the same category to be significantly related to each other. In other words, citation relations between journals within the same category should be relatively strong. By contrast, journals in different categories may be only weakly linked or may even be completely unrelated. Thus, the rationale of our approach can be summarized as follows: A journal should cite or be cited by journals within its own category with a high frequency in comparison with journals outside its category. Based on this basic principle, we define two criteria to identify journals with questionable classifications. One criterion is that if a journal has only a very small number of citation relations with other journals within its own category, then we believe the classification of the journal to be questionable. The other criterion is that if a journal has many citation relations with journals in a category to which the journal itself does not belong, then it seems likely that the journal incorrectly has not been assigned to this category.

In order to define the two criteria more formally, we first introduce the notion of the relatedness of a journal and a category. Let $n_{i,c}$ denote the number of citations between journal $i$ and journals in category $c$, counting both citations from journal $i$ to journals in category $c$ and citations from journals in category $c$ to journal $i$. Furthermore, let $t_i$ denote the total number of citations of journal $i$, counting both citations from journal $i$ to other journals and citations from other journals to journal $i$. Then, the relatedness of journal $i$ and category $c$ is defined as

$$r_{i,c} = \frac{n_{i,c}}{t_i}.$$

In the calculation of the relatedness $r_{i,c}$, only citations for which both the citing and the cited publication were published within the period of analysis (2010-2014 in our case; see Section 4) are considered. The direction of a citation is ignored, so no distinction is made between incoming and outgoing citations. Furthermore, journal self-citations, which are citations to earlier publications in the same journal, are excluded from the calculation of the relatedness $r_{i,c}$. This is because journal self-citations do not provide useful information for determining the relatedness of a journal and a category. Additionally, it should be noted that the sum over all categories $c$ of the number of citations between journal $i$ and journals in category $c$, that is $\sum_c n_{i,c}$, will typically be greater than $t_i$, the total number of citations of journal $i$. This is caused by the fact that WoS and Scopus often assign journals to more than one category.

Based on the notion of the relatedness of a journal and a category, the two criteria that we use in this paper to study the accuracy of a classification system can be expressed as follows:

**Criterion I.** A journal $i$ is assigned to a category $c$, but the number of citations between journal $i$ and category $c$ is relatively small, that is $r_{i,c} \leq \alpha$, with the threshold $\alpha$ equal to for instance 0.05, 0.1, or 0.2.

**Criterion II.** A journal $i$ is not assigned to a category $c$, but the number of citations between journal $i$ and category $c$ is relatively large, that is $r_{i,c} \geq \beta$, with the threshold $\beta$ equal to for instance 0.5, 0.6, 0.7, 0.8 or 0.9.

Criterion I is used to select journals that have weak connections with their assigned categories, while Criterion II is used to identify journals that are not assigned to categories with which they have strong connections. If a journal satisfies either of the two criteria, it can be concluded that its assignment to categories seems questionable.

One point is worth highlighting. It would be difficult to use our citation-based criteria to examine the classification of journals with a quite small number of citations, for instance $t_i < 100$. Our citation-based approach does not provide sufficient evidence to evaluate the classification of these journals. However, if we completely exclude journals with $t_i < 100$ from our analysis, this could affect the relatedness of the remaining journals and categories in an undesirable way. Let us take the WoS category ASIAN STUDIES as an example to further clarify this point. 61 journals are assigned to this category, of which there are 53 for which $t_i < 100$. If we completely exclude these 53 journals from our analysis, the relatedness of the eight remaining ASIAN STUDIES journals with the ASIAN STUDIES category will most likely be strongly reduced. It may then incorrectly seem as if these eight journals should not have been assigned to the ASIAN STUDIES category. To avoid this problem, we do not exclude any journals when determining the relatedness of journals and categories. However, because our citation-based criteria are not sufficiently reliable for journals with a small number of citations, we leave out these journals when presenting the results of our analysis.

## 4. Data

Our analysis is based on data from the in-house WoS and Scopus databases of the Centre for Science and Technology Studies (CWTS) at Leiden University. For WoS, journals in three citation indices are included, namely the Science Citation Index Expanded (SCIE), the Social Sciences Citation Index (SSCI), and the Arts & Humanities Citation Index (A&HCI). It should be noted that conference proceedings and books are excluded both in WoS and in Scopus; only journals and book series are included. For simplicity, in this paper the term 'journal' is used to refer both to journals and to book series. As explained in Subsection 2.3, WoS provides two classification systems, namely a system of categories and a system of research areas. Our focus in this paper is on the WoS categories classification system.

We retrieved from the WoS and Scopus databases all journals that have publications between 2010 and 2014.[1] This five-year time window was determined based on two considerations. On the one hand, journal classification systems are not entirely stable. The category assignment of a journal in WoS and Scopus sometimes changes, and in some cases entirely new categories are established in these classification systems. The longer the time window that is used, the more the analysis becomes sensitive to changes in a classification system. On the other hand, as indicated by Klavans and Boyack (2015), the use of direct citation relations requires a sufficiently long time widow. The choice of a five-year time window is a trade-off between on the one hand the stability of journal classification systems and on the other hand the accuracy of our approach based on direct citation relations.

During our five-year period of analysis, the producers of the databases have changed the category assignments of some journals. This was handled by taking the most recent category assignments of a journal. Table 1 shows some basic statistics on the classification systems of the two databases. In the case of Scopus, it should be noted that in the Scopus classification system, which consists of two levels, journals can be assigned both to categories at the top level and to categories at the bottom level. In Table 1, all category assignments in the Scopus classification system are counted, both at the top level and at the bottom level.

Table 1. Statistics on the assignment of journals to categories in WoS and Scopus

|  | WoS | Scopus |
|---|---|---|
| No. of publications | 9,124,596 | 10,770,432 |
| No. of journals | 12,393 | 24,015 |
| No. of categories | 251 | 331 |
| No. of journal-category assignments | 19,258 | 50,864 |
| Max. no. of categories per journal | 6 | 27 |

[1] In the case of WoS, journals that ceased publishing during the period 2010-2014 are not included in the analysis. This is because we do not have data on the category assignments of these journals.

| Avg. no. of categories per journal | 1.6 | 2.1 |
| --- | --- | --- |

As can be seen in Table 1, the number of Scopus journals included in the analysis is almost twice as large as the number of WoS journals, and Scopus also includes 80 more categories than WoS. Furthermore, although both databases often assign journals to multiple categories, we found that Scopus tends to assign journals to more categories than WoS. WoS assigns journals to at most six categories, whereas in Scopus there turns out to be a journal that is assigned to 27 categories.[2] Additionally, we found that the average number of categories to which journals belong equals 1.6 in WoS and 2.1 in Scopus. This shows that on average journals have significantly more category assignments in Scopus than in WoS. Figure 1 displays the distribution of journals in WoS and Scopus based on the number of categories to which they are assigned. As can be seen, almost 60% of all journals in WoS belong to only one category, whereas in Scopus more than 60% of all journals are assigned to two or more categories.

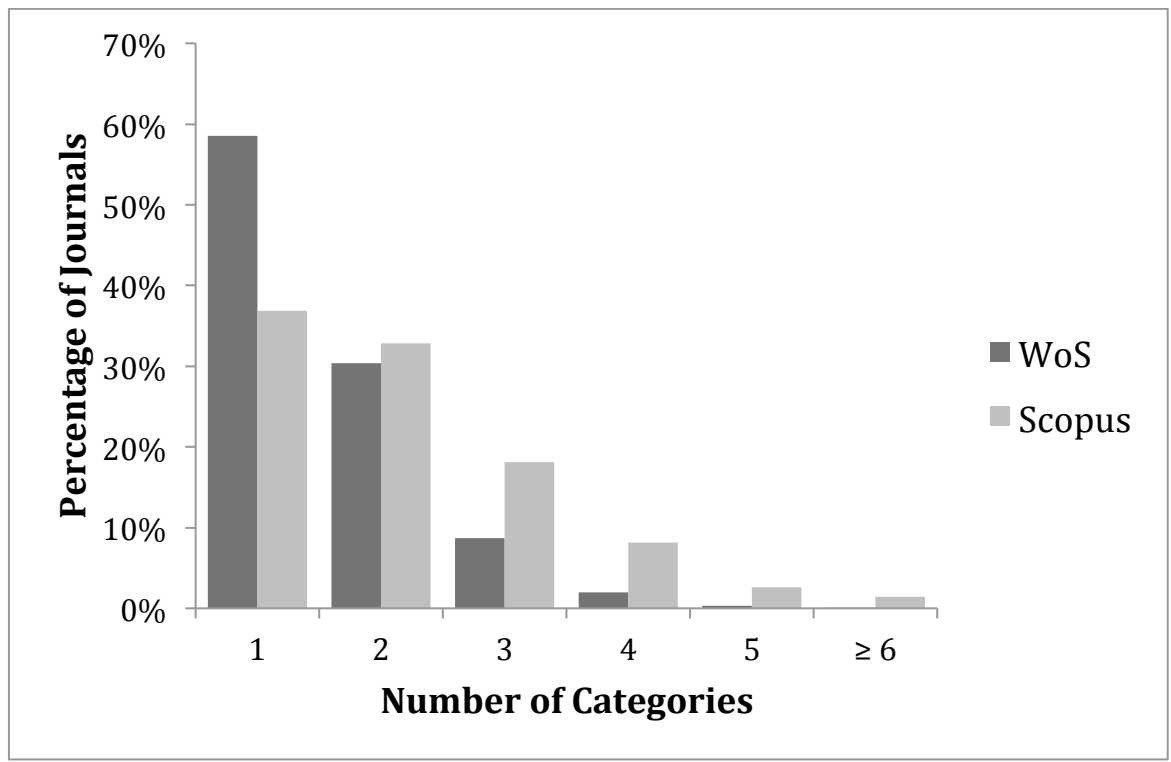

Figure 1. Distribution of journals in WoS and Scopus based on the number of categories to which they are assigned

## 5. Results

First of all, it should be mentioned that journals with $t_i < 100$ will not be included in the presentation of the results of our analysis. As discussed in Section 3, our citation-based approach may not be sufficiently reliable for these journals. WoS has 1,390 journals with $t_i < 100$, accounting for 11% of the total number of WoS journals, whereas Scopus has 5,808 journals with $t_i < 100$, which is 24% of the total.[3] Hence, Scopus has more journals with $t_i < 100$ than WoS not only in an absolute sense but also from a relative point of view. Taking a further look at Scopus journals with $t_i < 100$, it turns out that they can be roughly divided into three groups. One group consists of arts and humanities journals, another group consists of newly included journals, and a third group consists of non-English language journals. As already explained in Section 3, we emphasize that all journals, including those with $t_i < 100$, are taken into account in the calculation of the relatedness of journals and categories.

---

[2] The journal assigned to 27 categories is *Journal of Gambling Studies*. The journal with the second-largest number of category assignments in Scopus is *AMB Express*, which belongs to 16 categories.

The following subsections present the results of our analysis. Subsections 5.1 and 5.2 provide the results obtained using Criteria I and II, respectively. Subsection 5.3 reports the results obtained by combining Criteria I and II. Subsection 5.4 presents an in-depth analysis for the field of Library and Information Science. We note that detailed results of our analysis are available online.[4]

### 5.1. Criterion I: Journals assigned to a category with which they do not have a strong citation connection

A journal satisfies Criterion I if it is assigned to a certain category while the number of citations between the journal and other journals belonging to the same category is relatively small. More precisely, a journal $i$ satisfies Criterion I if it is assigned to a certain category $c$ even though $r_{i,c}$ is below a certain threshold $\alpha$. We use three values for the parameter $\alpha$. By using multiple parameter values, we get insight into the sensitivity of our results to the choice of the parameter value. One parameter value that we use is $\alpha = 0.05$. Using this parameter value, a journal satisfies Criterion I if the journal belongs to a category while the citations between the journal and other journals belonging to the same category account for less than 5% of the total number of citations of the journal. The other parameter values that we use are $\alpha = 0.1$ and $\alpha = 0.2$.

Before we present the results obtained using Criterion I, it should be noted that the classification systems of both WoS and Scopus include a number of special categories. WoS and Scopus both have a category that covers journals with a broad multidisciplinary scope, such as *Nature*, *Science*, and *PLoS ONE*.[5] Besides this, WoS also has a number of categories with words such as 'multidisciplinary', 'interdisciplinary', or 'general' in their label. Examples of these categories are AGRICULTURE, MULTIDISCIPLINARY and SOCIAL SCIENCES, INTERDISCIPLINARY. Likewise, in the case of Scopus, there are categories with 'miscellaneous' in their label.

Most categories in the classification systems of WoS and Scopus are intended to represent scientific fields, but this is not the case for the special categories discussed above. These special categories are not intended to represent scientific fields. However, Criterion I aims to test whether a journal belonging to a certain category is reasonably well connected, in terms of citations, to other journals belonging to the same category. This criterion is meaningful only if a category is intended to represent a scientific field. For categories that do not have such a function and that instead aim to cover a more heterogeneous or multidisciplinary set of journals, Criterion I is not meaningful. Because of this, we do not use Criterion I to examine the accuracy of assignments of journals to the above-discussed special categories. In the rest of this paper, we will refer to these special categories simply as multidisciplinary categories. In the case of WoS, there are 13 multidisciplinary categories, which are listed in Table A1 in the appendix. As already mentioned, the Scopus classification system has two levels and journals can be assigned to categories at both levels. The top-level Scopus categories are seen as multidisciplinary categories in this paper. In total, Scopus has 53 multidisciplinary categories. These are the category MULTIDISCIPLINARY and the 26 other top-level categories and within each of these 26 categories a bottom-level miscellaneous category such as MATHEMATICS (MISCELLANEOUS) or MEDICINE (MISCELLANEOUS).

---

[3] In the case of WoS, the journals with $t_i < 100$ contain 423,364 publications, accounting for 5% of the total number of WoS publications, whereas the Scopus journals with $t_i < 100$ contain 623,346 publications, which is 6% of the total.

[4] The results are available at <u>www.ludowaltman.nl/wos_scopus/</u>. On this webpage, extensive statistics on the relatedness of journals and categories are provided, both for WoS and for Scopus.

[5] This category is labeled MULTIDISCIPLINARY SCIENCES in WoS, whereas it is labeled MULTIDISCIPLINARY in Scopus.

Table 2 provides some basic statistics on the assignment of journals to categories in WoS and Scopus when journals with $t_i < 100$ and assignments of journals to multidisciplinary categories are excluded. The table shows the number of journals that belong to at least one non-multidisciplinary category and the number of assignments of journals to non-multidisciplinary categories. As can be seen in the table, in the case of Scopus the constraints that we have introduced cause a much larger decrease in the number of journals and the number of journal-category assignments than in the case of WoS.

Table 2. Statistics on the assignment of journals to categories in WoS and Scopus (excluding journals with $t_i < 100$ and excluding assignments to multidisciplinary categories)

|  | WoS | Scopus |
| --- | --- | --- |
| No. of publications | 7,835,836 | 8,299,765 |
| % of all publications | 86% | 77% |
| No. of journals | 10,386 | 15,934 |
| % of all journals | 84% | 66% |
| No. of journal-category assignments | 16,097 | 33,400 |
| % of all journal-category assignments | 84% | 66% |

Table 3 reports for both WoS and Scopus and for three values of the threshold $\alpha$ the number of journals and the number of journal-category assignments that satisfy Criterion I. A journal satisfies Criterion I if at least one of its category assignments satisfies the criterion. As can be seen, both databases have assigned a significant number of journals to categories that according to Criterion I seem to be inappropriate. As can be expected, the number of journals and journal-category assignments satisfying Criterion I increases as the threshold $\alpha$ increases. Moreover, no matter which threshold $\alpha$ is considered, Scopus performs substantially worse than WoS, not only in the absolute number of journals and journal-category assignments satisfying Criterion I but, more importantly, also in the percentage of journals and journal-category assignments satisfying the criterion. For $\alpha = 0.05$ and $\alpha = 0.1$, the percentage of journals and journal-category assignments satisfying Criterion I is more than two times higher for Scopus than for WoS. Nevertheless, even in the case of WoS, for $\alpha = 0.1$ we still find that 16% of the journals have one or more questionable category assignments.

Table 3. Summary of the results from Criterion I (excluding journals with $t_i < 100$ and excluding assignments to multidisciplinary categories)

| Threshold $\alpha$ | WoS | | Scopus | |
| --- | --- | --- | --- | --- |
|  | No. of journals (% of all journals) | No. of journal-category assignments (% of all journal-category assignments) | No. of journals (% of all journals) | No. of journal-category assignments (% of all journal-category assignments) |

| 0.05 | 762 (7%) | 838 (5%) | 3,314 (21%) | 4,407 (13%) |
|------|----------|----------|-------------|--------------|
| 0.10 | 1,683 (16%) | 1,947 (12%) | 5,653 (35%) | 8,500 (25%) |
| 0.20 | 3,623 (35%) | 4,795 (30%) | 8,939 (56%) | 15,751 (47%) |

Next, we identify WoS and Scopus categories with a high percentage of journals satisfying Criterion I. The identified categories may be seen as the most problematic categories in the two databases, because many of the journals belonging to these categories are only weakly connected to each other in terms of citations. We select categories that include at least 10 journals with $t_i \geq 100$ and that, for $\alpha = 0.1$, have at least 50% of their journals satisfying Criterion I. The results for WoS and Scopus are reported in Tables 4 and 5, respectively. In the case of WoS 17 categories have been identified, whereas in the case of Scopus 76 categories have been identified, so more than four times as many as in the case of WoS. There are three categories that have been identified in the case of both databases: ARCHITECTURE, BIOPHYSICS, and MEDICAL LABORATORY TECHNOLOGY.

Table 4. Categories in which at least 50% of the journals satisfy Criterion I (WoS; $\alpha = 0.1$; excluding journals with $t_i < 100$)

| WoS category | No. of journals | No. of journals with $r_{i,c} \leq 0.1$ | % of journals with $r_{i,c} \leq 0.1$ |
|--------------|-----------------|------------------------------------------|----------------------------------------|
| MEDICINE, RESEARCH & EXPERIMENTAL | 121 | 104 | 86% |
| ARCHITECTURE | 11 | 9 | 82% |
| BIOLOGY | 83 | 66 | 80% |
| SOCIAL ISSUES | 36 | 28 | 78% |
| MATERIALS SCIENCE, CHARACTERIZATION & TESTING | 33 | 24 | 73% |
| MICROSCOPY | 10 | 7 | 70% |
| MEDICAL LABORATORY TECHNOLOGY | 28 | 19 | 68% |
| ANATOMY & MORPHOLOGY | 19 | 13 | 68% |
| BIOPHYSICS | 69 | 43 | 62% |
| CULTURAL STUDIES | 28 | 17 | 61% |
| FILM, RADIO, TELEVISION | 10 | 6 | 60% |
| COMPUTER SCIENCE, CYBERNETICS | 22 | 13 | 59% |
| CHEMISTRY, APPLIED | 67 | 39 | 58% |
| ETHNIC STUDIES | 14 | 8 | 57% |
| PRIMARY HEALTH CARE | 18 | 10 | 56% |
| PHYSIOLOGY | 82 | 45 | 55% |
| PSYCHOLOGY, BIOLOGICAL | 15 | 8 | 53% |

Table 5. Categories in which at least 50% of the journals satisfy Criterion I (Scopus; $\alpha = 0.1$; excluding journals with $t_i < 100$)

| Scopus category | No. of journals | No. of journals with $r_{i,c} \leq 0.1$ | % of journals with $r_{i,c} \leq 0.1$ |
|-----------------|-----------------|------------------------------------------|----------------------------------------|
| LIFE-SPAN AND LIFE-COURSE STUDIES | 38 | 38 | 100% |
| COMMUNITY AND HOME CARE | 33 | 33 | 100% |
| MEDICAL LABORATORY TECHNOLOGY | 30 | 30 | 100% |
| EMBRYOLOGY | 17 | 17 | 100% |
| RESEARCH AND THEORY | 10 | 10 | 100% |
| DEVELOPMENTAL NEUROSCIENCE | 30 | 29 | 97% |
| ADVANCED AND SPECIALIZED NURSING | 44 | 42 | 95% |
| PEDIATRICS | 22 | 21 | 95% |
| ECOLOGICAL MODELING | 20 | 19 | 95% |
| INDUSTRIAL RELATIONS | 33 | 30 | 91% |
| ENDOCRINE AND AUTONOMIC SYSTEMS | 22 | 20 | 91% |
| COMPUTATIONAL MECHANICS | 32 | 29 | 91% |
| MEDICAL AND SURGICAL NURSING | 20 | 18 | 90% |
| CONSERVATION | 10 | 9 | 90% |
| COMPLEMENTARY AND MANUAL THERAPY | 10 | 9 | 90% |
| ARCHITECTURE | 33 | 29 | 88% |
| SAFETY RESEARCH | 42 | 36 | 86% |
| HISTOLOGY | 55 | 47 | 85% |
| BIOCHEMISTRY (MEDICAL) | 56 | 47 | 84% |
| HEALTH INFORMATION MANAGEMENT | 18 | 15 | 83% |
| ECONOMIC GEOLOGY | 20 | 16 | 80% |
| PROCESS CHEMISTRY AND TECHNOLOGY | 29 | 23 | 79% |
| PSYCHIATRIC MENTAL HEALTH | 38 | 30 | 79% |
| STRUCTURAL BIOLOGY | 46 | 36 | 78% |
| FAMILY PRACTICE | 32 | 25 | 78% |
| BIOPHYSICS | 119 | 90 | 76% |
| COMPUTATIONAL THEORY AND MATHEMATICS | 96 | 72 | 75% |
| EMERGENCY NURSING | 20 | 15 | 75% |
| MANAGEMENT INFORMATION SYSTEMS | 64 | 47 | 73% |
| FUNDAMENTALS AND SKILLS | 15 | 11 | 73% |
| RADIATION | 40 | 29 | 73% |
| RADIOLOGICAL AND ULTRASOUND TECHNOLOGY | 43 | 31 | 72% |
| NUMERICAL ANALYSIS | 39 | 28 | 72% |
| CLINICAL BIOCHEMISTRY | 122 | 87 | 71% |
| MODELING AND SIMULATION | 193 | 136 | 70% |
| HUMAN FACTORS AND ERGONOMICS | 27 | 19 | 70% |
| COLLOID AND SURFACE CHEMISTRY | 13 | 9 | 69% |
| BEHAVIORAL NEUROSCIENCE | 61 | 42 | 69% |
| INSTRUMENTATION | 80 | 55 | 69% |
| ANATOMY | 38 | 26 | 68% |
| GLOBAL AND PLANETARY CHANGE | 46 | 31 | 67% |
| MOLECULAR MEDICINE | 162 | 108 | 67% |
| STRATIGRAPHY | 33 | 22 | 67% |
| PHARMACY | 21 | 14 | 67% |
| EPIDEMIOLOGY | 85 | 55 | 65% |
| FLUID FLOW AND TRANSFER PROCESSES | 39 | 25 | 64% |
| MEDIA TECHNOLOGY | 33 | 21 | 64% |
| CONTROL AND OPTIMIZATION | 50 | 31 | 62% |
| PHYSIOLOGY (MEDICAL) | 93 | 57 | 61% |
| VISUAL ARTS AND PERFORMING ARTS | 50 | 30 | 60% |
| LEADERSHIP AND MANAGEMENT | 30 | 18 | 60% |
| AGING | 31 | 18 | 58% |
| NEUROPSYCHOLOGY AND PHYSIOLOGICAL PSYCHOLOGY | 57 | 33 | 58% |
| HUMAN-COMPUTER INTERACTION | 71 | 41 | 58% |
| COMPUTER GRAPHICS AND COMPUTER-AIDED DESIGN | 52 | 30 | 58% |
| COMPUTATIONAL MATHEMATICS | 96 | 55 | 57% |
| FOOD ANIMALS | 28 | 16 | 57% |
| SAFETY, RISK, RELIABILITY AND QUALITY | 100 | 57 | 57% |
| INFORMATION SYSTEMS AND MANAGEMENT | 63 | 35 | 56% |
| BIOMATERIALS | 63 | 35 | 56% |
| MATERNITY AND MIDWIFERY | 20 | 11 | 55% |
| ONCOLOGY (NURSING) | 15 | 8 | 53% |
| ISSUES, ETHICS AND LEGAL ASPECTS | 36 | 19 | 53% |
| OCEAN ENGINEERING | 55 | 29 | 53% |
| HISTORY AND PHILOSOPHY OF SCIENCE | 80 | 42 | 53% |
| COMPUTERS IN EARTH SCIENCES | 21 | 11 | 52% |
| SIGNAL PROCESSING | 65 | 34 | 52% |
| DEVELOPMENT | 158 | 82 | 52% |
| BIOLOGICAL PSYCHIATRY | 35 | 18 | 51% |
| MATHEMATICAL PHYSICS | 43 | 22 | 51% |
| HEALTH (SOCIAL SCIENCE) | 200 | 102 | 51% |
| CELLULAR AND MOLECULAR NEUROSCIENCE | 81 | 41 | 51% |
| BIOTECHNOLOGY | 228 | 114 | 50% |
| DEVELOPMENTAL BIOLOGY | 78 | 39 | 50% |
| CRITICAL CARE NURSING | 18 | 9 | 50% |
| BIOENGINEERING | 127 | 63 | 50% |

*5.2. Criterion II: Journals not assigned to a category with which they have a strong citation connection*

A journal satisfies Criterion II if it is not assigned to a certain category while the number of citations between the journal and other journals that do belong to the category is relatively large. More precisely, a journal $i$ satisfies Criterion II if it is not assigned to a certain category $c$ even though $r_{i,c}$ is above a certain threshold $\beta$. Like in the previous section, we use multiple parameter values. Five different values are used for the parameter $\beta$.

Table 6 presents for both WoS and Scopus and for five values of the threshold $\beta$ the number of journals that satisfy Criterion II.[6] As can be expected, as the threshold $\beta$ increases, the number of journals satisfying Criterion II decreases. For $\beta$ = 0.9, there is no WoS journal satisfying Criterion II and there are only two Scopus journals satisfying the criterion. Even for $\beta$ = 0.5, less than 5% of all journals in WoS and Scopus satisfy Criterion II. Hence, it turns out that according to Criterion II both databases perform reasonably well. Both databases sometimes do not assign a journal to a category even though in terms of citations the journal is strongly connected to the category, but this happens only in a relatively limited number of cases. Looking at the percentages reported in Table 6, it can be seen that WoS performs somewhat better than Scopus.

---

[6] In exceptional cases, a journal may have multiple categories for which it satisfies Criterion II. In that case, the journal is counted only once in Table 6.

Table 6. Summary of the results from Criterion II (excluding journals with $t_i < 100$)

| Threshold $\beta$ | WoS | | Scopus | |
|---|---|---|---|---|
| | No. of journals | % of all journals | No. of journals | % of all journals |
| 0.5 | 236 | 2.14% | 722 | 3.97% |
| 0.6 | 87 | 0.79% | 259 | 1.42% |
| 0.7 | 27 | 0.25% | 82 | 0.45% |
| 0.8 | 4 | 0.04% | 25 | 0.14% |
| 0.9 | 0 | 0.00% | 2 | 0.01% |

For each database, we further identify categories for which there are at least 10 journals that are not assigned to the category but that according to Criterion II, with $\beta = 0.6$, should be assigned to it. The results for WoS and Scopus are presented in Tables 7 and 8, respectively. In both tables, the first column lists the categories to which journals should have been assigned according to Criterion II, but to which they are not assigned. Comparing the two tables, we note that some similarities can be observed. The categories ECONOMICS and ENGINEERING, ELECTRICAL & ELECTRONIC in Table 7 are similar to the categories ECONOMICS AND ECONOMETRICS and ELECTRICAL AND ELECTRONIC ENGINEERING in Table 8.

Table 7. Categories for which there are at least 10 journals that are not assigned to the category but that according to Criterion II should be assigned to it (WoS; $\beta = 0.6$; excluding journals with $t_i < 100$)

| WoS category | No. of journals | No. of journals with $r_{i,c} \geq 0.6$ not assigned to category |
|---|---|---|
| ECONOMICS | 324 | 15 |
| MATHEMATICS, APPLIED | 254 | 11 |
| ENGINEERING, ELECTRICAL & ELECTRONIC | 243 | 10 |

Table 8. Categories for which there are at least 10 journals that are not assigned to the category but that according to Criterion II should be assigned to it (Scopus; $\beta = 0.6$; excluding journals with $t_i < 100$)

| Scopus category | No. of journals | No. of journals with $r_{i,c} \geq 0.6$ not assigned to category |
|---|---|---|
| ECONOMICS AND ECONOMETRICS | 455 | 55 |
| ECOLOGY, EVOLUTION, BEHAVIOR AND SYSTEMATICS | 496 | 24 |
| SOCIOLOGY AND POLITICAL SCIENCE | 643 | 15 |
| SPACE AND PLANETARY SCIENCE | 70 | 12 |
| EDUCATION | 754 | 12 |
| ELECTRICAL AND ELECTRONIC ENGINEERING | 521 | 12 |
| THEORETICAL COMPUTER SCIENCE | 104 | 11 |

*5.3. Combining Criteria I and II: Journals with the most questionable category assignments*

We now combine Criteria I and II to examine the journals with the most questionable category assignments in WoS and Scopus. A journal satisfies both Criterion I and Criterion II if on the one hand it has weak connections, in terms of citations, with its assigned categories while on the other hand it has a strong connection with a category to which it is not assigned. More precisely, our focus is on journals for which the current category assignments all satisfy Criterion I, while there is an alternative category assignment that satisfies Criterion II. For these journals, we can conclude that their assignment to categories is even more questionable than for a journal that satisfies only one of the two criteria. The results discussed below are obtained using the parameter values $\alpha = 0.1$ and $\beta = 0.6$.

In WoS, there is only one journal that satisfies the combined Criteria I and II, namely *Australian Journal of Management*. This journal belongs to the category MANAGEMENT, even though its relatedness with this category is only 0.07. However, *Australian Journal of Management* is actually strongly connected with the category BUSINESS, FINANCE, with a relatedness of 0.74. The aims and scope statement of the journal is as follows:

> The objectives of the *Australian Journal of Management* are to encourage and publish research in the field of management … Consistent with the policy, the Australian Journal of Management publishes peer-reviewed research in accounting, applied economics, finance, industrial relations, political science, psychology, statistics, and other disciplines. This is providing that the application is to management and research in areas such as marketing, corporate strategy, operations management, organisation development, decision analysis, and other problem-focused paradigms.[7]

In the aims and scope statement, several fields such as management, accounting, applied economics, finance, etc. are mentioned. However, taking a further look at journals related to *Australian Journal of Management*, it turns out that none of the ten journals that have most citation relations with *Australian Journal of Management* belong to the category MANAGEMENT. Instead, nine of these journals are assigned to the category BUSINESS, FINANCE. It seems that WoS has classified *Australian Journal of Management* based on its title and perhaps also its aims and scope statement; however, from a citation perspective, the classification of this journal should be reconsidered.

In Scopus, there are 32 journals that satisfy the combined Criteria I and II. The list of journals is shown in Table A2 in the appendix. We now discuss two journals in more detail.

Like in the case of WoS, we consider a journal with an assignment to a management-related category, namely *Cooperation and Conflict*. In Scopus, this journal is assigned to the category STRATEGY AND MANAGEMENT. However, it turns out that the journal has an extremely weak connection with this category, with a relatedness of 0.01; conversely, the journal is strongly connected with the category POLITICAL SCIENCE AND INTERNATIONAL RELATIONS, with a relatedness of 0.67. *Cooperation and Conflict* states its scope in a very explicit way: "the aim of *Cooperation and Conflict* is to promote research on and understanding of international relations"[8]. This statement is in full agreement with the results obtained by taking a citation perspective, and it contradicts the category assignment of the journal in Scopus.

As a second example, we take *Mobilization*, which is a journal assigned to the category TRANSPORTATION in Scopus. It turns out that the journal has no citation relations at all with this category (relatedness of 0.00), while it has a strong connection in terms of citations with the category SOCIOLOGY AND POLITICAL SCIENCE (relatedness of 0.64). The journal summarizes its scope as follows:

> *Mobilization* is a review of research about social and political movements, strikes, riots, protests, insurgencies, revolutions, and other forms of contentious politics. Its goal is to

---

advance the systematic, scholarly, and scientific study of these phenomena, and to provide a forum for the discussion of methodologies, theories, and conceptual approaches across the disciplines of sociology, political science, social psychology, and anthropology.[9]

Based on this statement, it is clear that *Mobilization* should be assigned to the category SOCIOLOGY AND POLITICAL SCIENCE instead of the category TRANSPORTATION, which confirms our citation-based findings. The examples of *Cooperation and Conflict* and *Mobilization* also provide evidence that our citation-based criteria give useful indications of misclassified journals.

Based on the three journals discussed above, we conclude that journals satisfying the combined Criteria I and II can be classified into at least two types. One type refers to journals for which there is a discrepancy between on the one hand their title and their scope statement and on the other hand what they have actually published. *Australian Journal of Management* is an example of such a journal. Based on its scope statement, its WoS category assignment seems reasonable, but the scope statement itself may not be fully accurate. The second type refers to journals that seem to have been assigned to a category based only on their title. An example is *Mobilization*. The title of this journal seems to have been misinterpreted and the scope statement seems to have been ignored, leading to an incorrect category assignment in Scopus.

*5.4. In-depth analysis for the field of Library and Information Science*

In this subsection, we take the field of Library and Information Science (LIS) as an example to conduct a more in-depth analysis. We choose to focus on the LIS field because many readers of this paper are likely to be familiar with this field. The analysis that we present can also be helpful to examine whether the criteria that we use to identify journals with questionable category assignments are appropriate and whether they yield meaningful results. In WoS the LIS field is represented by the category INFORMATION SCIENCE & LIBRARY SCIENCE, whereas it is represented by the category LIBRARY & INFORMATION SCIENCES in Scopus.

WoS and Scopus have respectively 85 and 209 LIS journals. The differences in journal coverage between the WoS and Scopus LIS categories are shown in Table 9. As can be seen, there are 54 journals that are assigned to the LIS category both in WoS and in Scopus. However, there are also a substantial number of journals that are included in both databases but that belong to the LIS category in only one of the databases. This finding is in accordance with the study by Abrizah et al. (2013), who also pointed out differences in journal coverage between the LIS categories in WoS and Scopus. Of the 85 and 209 LIS journals in WoS and Scopus, there are respectively 75 and 143 with $t_i \geq 100$. In the rest of this subsection, results are presented only for these journals.

Table 9. Comparison of LIS journals in WoS and Scopus

| WoS | Scopus |
| --- | --- |
| Total number of LIS journals: 85 | Total number of LIS journals: 209 |
| In Scopus LIS category: 54 | In WoS LIS category: 54 |
| In Scopus, but not in LIS category: 24 | In WoS, but not in LIS category: 19 |
| Not in Scopus: 7 | Not in WoS: 136 |

---

[9] More detailed information on *Mobilization* is available at http://www.mobilization.sdsu.edu.

We first examine the assignment of journals to the WoS and Scopus LIS categories from the point of view of Criterion I. More specifically, we identify journals in WoS and Scopus that belong to the LIS category while the citations between the journal and other journals belonging to the LIS category account for less than 10% of the total number of citations of the journal. So we apply Criterion I using the parameter value $\alpha = 0.1$. Tables 10 and 11 report for WoS and Scopus the journals with an assignment to the LIS category that satisfies Criterion I. WoS has eight LIS journals (11% of the total number of LIS journals in WoS) satisfying Criterion I, whereas Scopus has 29 LIS journals (20%) satisfying the criterion. There are four journals (indicated in bold in Tables 10 and 11) that satisfy Criterion I in both databases. We note that some journals (e.g., *Information Systems Research*) are assigned to the LIS category in both databases but satisfy Criterion I in only one of the two databases.

Table 10. LIS journals satisfying Criterion I (WoS; $\alpha = 0.1$; excluding journals with $t_i < 100$)

| WoS journal | $n_{i,c}$ | $r_{i,c}$ |
| --- | --- | --- |
| ***Ethics and Information Technology*** | **36** | **0.10** |
| *Information Technology & Management* | 69 | 0.07 |
| *International Journal of Computer-Supported Collaborative Learning* | 15 | 0.03 |
| ***International Journal of Geographical Information Science*** | **14** | **0.01** |
| ***Journal of Health Communication*** | **62** | **0.02** |
| *Journal of the American Medical Informatics Association* | 75 | 0.01 |
| *Scientist* | 2 | 0.00 |
| ***Social Science Information sur les Sciences Sociales*** | **24** | **0.08** |

Table 11. LIS journals satisfying Criterion I (Scopus; $\alpha = 0.1$; excluding journals with $t_i < 100$)

| Scopus journal | $n_{i,c}$ | $r_{i,c}$ |
| --- | --- | --- |
| *Accountability in Research* | 10 | 0.02 |
| *Campus-Wide Information Systems* | 27 | 0.05 |
| *Canadian Journal of Program Evaluation* | 0 | 0.00 |
| *Computers in the Schools* | 6 | 0.02 |
| *Cuadernos.info* | 12 | 0.09 |
| *Development and Learning in Organisations* | 7 | 0.04 |
| *Education and Information Technologies* | 6 | 0.02 |
| ***Ethics and Information Technology*** | **27** | **0.05** |
| *IEEE Transactions on Information Theory* | 58 | 0.01 |
| *Information Communication and Society* | 123 | 0.06 |
| *Information Management and Computer Security* | 28 | 0.08 |
| *Information Retrieval* | 46 | 0.09 |
| *Information Systems Research* | 175 | 0.07 |
| *Intelligent Systems Reference Library* | 20 | 0.01 |
| *International Journal of Data Mining and Bioinformatics* | 6 | 0.01 |
| ***International Journal of Geographical Information Science*** | **24** | **0.01** |
| *International Journal of Law and Information Technology* | 7 | 0.05 |
| *Journal of Chemical Information and Modeling* | 385 | 0.02 |
| *Journal of Classification* | 0 | 0.00 |
| *Journal of Digital Information Management* | 10 | 0.03 |
| ***Journal of Health Communication*** | **43** | **0.01** |
| *Journal of Information and Computational Science* | 61 | 0.01 |
| *Journal of Information Science and Engineering* | 22 | 0.02 |
| *Knowledge Management Research and Practice* | 46 | 0.07 |
| *Language Resources and Evaluation* | 9 | 0.03 |
| *Lecture Notes in Control and Information Sciences* | 20 | 0.02 |
| *Notes and Queries* | 2 | 0.02 |
| *Social Science Computer Review* | 132 | 0.09 |
| ***Social Science Information*** | **21** | **0.05** |

We now turn to Criterion II, so we identify journals that are not assigned to the LIS category while the number of citations between the journal and journals belonging to the LIS category is relatively large. We use the parameter value $\beta = 0.6$. In the case of WoS, there turn out to be no journals that satisfy Criterion II. In the case of Scopus, there are three journals that satisfy the criterion. These journals are *Portal: Libraries and the Academy*, which is assigned to the categories COMPUTER SCIENCE APPLICATIONS and INFORMATION SYSTEMS, *Online (Wilton, Connecticut)*, which belongs to the category DEVELOPMENT, and *Public Services Quarterly*, which is assigned to the categories ACCOUNTING and PUBLIC ADMINISTRATION. These three journals do not belong to the LIS category in Scopus even though for each of these journals citations between the journal and journals belonging to the LIS category account for more than 60% of the total number of citations of the journal.

Researchers in the field of bibliometrics and scientometrics may also expect *Journal of Informetrics* to satisfy Criterion II in the case of Scopus. *Journal of Informetrics* is focused strongly on bibliometric and scientometric studies, but it is not assigned to the LIS category in Scopus (unlike for instance *Scientometrics*, which does belong to the LIS category). Taking a further look at this specific case, it turns out that *Journal of Informetrics* has a relatively strong connection with the LIS category, with a relatedness of 0.45. Although it does not satisfy Criterion II for $\beta = 0.6$, we still find that LIS is the category with which *Journal of Informetrics* has the strongest connection in terms of citations. In fact, the relatedness of *Journal of Informetrics* with the LIS category turns out to be higher than the total relatedness of the journal with the five categories to which it is assigned in Scopus (i.e., APPLIED MATHEMATICS, COMPUTER SCIENCE APPLICATIONS, MANAGEMENT SCIENCE & OPERATIONS RESEARCH, MODELING & SIMULATION, and STATISTICS & PROBABILITY).

We are also interested in exploring the accuracy of other category assignments of LIS journals. For instance, *Scientometrics* is assigned not only to the LIS category in Scopus but also to the category LAW. We aim to examine whether the assignment of *Scientometrics* to the category LAW seems justified. We use Criterion I to identify LIS journals that have weak connections with other categories to which they are assigned. Like above, we use the parameter value $\alpha = 0.1$. The results are shown in Tables 12 and 13. It turns out that WoS has five LIS journals with questionable assignments to other categories, whereas Scopus has 38 LIS journals with assignments to other categories that seem questionable. There are three LIS journals that have questionable category assignments in both databases (indicated in bold in Tables 12 and 13), namely *International Journal of Geographical Information Science*, *Ethics and Information Technology*, and *Journal of Health Communication*.

Table 12. Assignments of LIS journals to other categories satisfying Criterion I (WoS; $\alpha = 0.1$; excluding journals with $t_i < 100$)

| WoS journal | $n_{i,c}$ | $r_{i,c}$ | WoS category |
|---|---|---|---|
| ***Ethics and Information Technology*** | **33** | **0.09** | **ETHICS** |
| ***International Journal of Geographical Information Science*** | **101** | **0.03** | **COMPUTER SCIENCE, INFORMATION SYSTEMS** |
| ***Journal of Health Communication*** | **351** | **0.10** | **COMMUNICATION** |
| *Journal of the American Medical Informatics Association* | 538 | 0.07 | COMPUTER SCIENCE, INFORMATION SYSTEMS |
| *Telecommunications Policy* | 94 | 0.08 | COMMUNICATION |
| *Telecommunications Policy* | 107 | 0.09 | TELECOMMUNICATIONS |

Table 13. Assignments of LIS journals to other categories satisfying Criterion I (Scopus; $\alpha = 0.1$; excluding journals with $t_i < 100$)

| Scopus journal | $n_{i,c}$ | $r_{i,c}$ | Scopus category |
|---|---|---|---|
| *Accountability in Research* | 55 | 0.10 | EDUCATION |
| *Campus-Wide Information Systems* | 24 | 0.05 | COMPUTER NETWORKS AND COMMUNICATIONS |
| *Collection Management* | 2 | 0.00 | STRATEGY AND MANAGEMENT |
| ***Ethics and Information Technology*** | **42** | **0.08** | **COMPUTER SCIENCE APPLICATIONS** |
| *Government Information Quarterly* | 134 | 0.06 | LAW |
| *Health Information and Libraries Journal* | 6 | 0.01 | HEALTH INFORMATION MANAGEMENT |
| *IEEE Transactions on Information Theory* | 483 | 0.04 | INFORMATION SYSTEMS |
| *Information Management and Computer Security* | 27 | 0.07 | BUSINESS AND INTERNATIONAL MANAGEMENT |
| *Information Management and Computer Security* | 14 | 0.04 | MANAGEMENT SCIENCE AND OPERATIONS RESEARCH |
| *Information Processing and Management* | 71 | 0.04 | MANAGEMENT SCIENCE AND OPERATIONS RESEARCH |
| *Information Processing and Management* | 23 | 0.01 | MEDIA TECHNOLOGY |
| *Information Resources Management Journal* | 22 | 0.09 | STRATEGY AND MANAGEMENT |
| *Information Systems Research* | 153 | 0.06 | COMPUTER NETWORKS AND COMMUNICATIONS |
| *Intelligent Systems Reference Library* | 125 | 0.06 | INFORMATION SYSTEMS AND MANAGEMENT |
| *International Journal of Data Mining and Bioinformatics* | 23 | 0.04 | INFORMATION SYSTEMS |
| ***International Journal of Geographical Information Science*** | **228** | **0.06** | **INFORMATION SYSTEMS** |
| *International Journal of Information Management* | 231 | 0.07 | COMPUTER NETWORKS AND COMMUNICATIONS |
| *International Journal of Information Science and Management* | 14 | 0.06 | INFORMATION SYSTEMS AND MANAGEMENT |
| *International Journal of Information Science and Management* | 17 | 0.08 | MANAGEMENT INFORMATION SYSTEMS |
| *Journal of Business and Finance Librarianship* | 1 | 0.01 | MANAGEMENT INFORMATION SYSTEMS |
| *Journal of Business and Finance Librarianship* | 4 | 0.02 | MARKETING |
| *Journal of Cheminformatics* | 22 | 0.01 | COMPUTER GRAPHICS AND COMPUTER-AIDED DESIGN |
| *Journal of Cheminformatics* | 270 | 0.10 | PHYSICAL AND THEORETICAL CHEMISTRY |
| *Journal of Digital Information Management* | 10 | 0.03 | MANAGEMENT INFORMATION SYSTEMS |
| *Journal of Educational Media and Library Science* | 1 | 0.01 | CONSERVATION |
| *Journal of Electronic Resources in Medical Libraries* | 2 | 0.01 | HEALTH (SOCIAL SCIENCE) |
| ***Journal of Health Communication*** | **397** | **0.09** | **COMMUNICATION** |
| *Journal of Information and Computational Science* | 227 | 0.04 | COMPUTATIONAL THEORY AND MATHEMATICS |
| *Journal of Information and Computational Science* | 218 | 0.04 | COMPUTER GRAPHICS AND COMPUTER-AIDED DESIGN |
| *Journal of Information and Knowledge Management* | 32 | 0.08 | COMPUTER NETWORKS AND COMMUNICATIONS |
| *Journal of Information Science and Engineering* | 52 | 0.06 | COMPUTATIONAL THEORY AND MATHEMATICS |
| *Journal of Information Science and Engineering* | 87 | 0.10 | HARDWARE AND ARCHITECTURE |
| *Journal of Information Science and Engineering* | 28 | 0.03 | HUMAN-COMPUTER INTERACTION |
| *Journal of Library Administration* | 12 | 0.01 | PUBLIC ADMINISTRATION |
| *Journal of the Association for Information Science and Technology* | 24 | 0.03 | INFORMATION SYSTEMS AND MANAGEMENT |
| *Journal of Web Librarianship* | 38 | 0.08 | COMPUTER SCIENCE APPLICATIONS |
| *Knowledge Management Research and Practice* | 39 | 0.06 | MANAGEMENT INFORMATION SYSTEMS |
| *Language Resources and Evaluation* | 3 | 0.01 | EDUCATION |
| *OCLC Systems and Services* | 12 | 0.05 | EDUCATION |
| *Records Management Journal* | 1 | 0.01 | MANAGEMENT INFORMATION SYSTEMS |
| *Research Evaluation* | 59 | 0.06 | EDUCATION |
| *Scientometrics* | 168 | 0.02 | LAW |
| *Social Science Computer Review* | 105 | 0.08 | COMPUTER SCIENCE APPLICATIONS |
| *Social Science Computer Review* | 100 | 0.08 | LAW |
| *Technical Services Quarterly* | 15 | 0.05 | COMPUTER SCIENCE APPLICATIONS |
| *World Patent Information* | 6 | 0.02 | BIOENGINEERING |
| *World Patent Information* | 4 | 0.01 | RENEWABLE ENERGY, SUSTAINABILITY AND THE ENVIRONMENT |

## 6. Discussion and Conclusions

This study examined and compared the accuracy of the WoS and Scopus journal classification systems. Based on direct citation relations between journals and categories, we defined two criteria to examine the category assignments of journals. Criterion I was used to identify journals that in terms of citations have weak connections with their assigned categories, and Criterion II was used to identify journals that are not assigned to categories with which they have strong connections. If a journal satisfies either of these two criteria, it can be concluded that the classification of the journal is questionable. Furthermore, we also used the combined Criteria I and II to identify journals that have weak connections with all their assigned categories while they have a strong connection with a category to which they are not assigned. These can be seen as the journals with the most questionable classification.

### 6.1. Research findings

Our most important findings regarding the accuracy of the WoS and Scopus journal classification systems can be summarized as follows. First, WoS performs much better than Scopus according to Criterion I. Using the parameter values $\alpha = 0.05$ and $\alpha = 0.1$, the percentage of journals and journal-category assignments satisfying Criterion I is more than two times higher for Scopus than for WoS. Hence, in Scopus journals are assigned to categories with which they are only weakly connected much more frequently than in WoS. Second, based on Criterion II, WoS and Scopus both perform reasonably well, with WoS having a somewhat better performance than Scopus. For all parameter values that were considered, less than 5% of all journals in WoS and Scopus satisfy Criterion II. In other words, if a journal is strongly connected to a category, WoS and Scopus typically assign the journal to that category. Third, WoS also presents a significantly better result than Scopus based on the combined Criteria I and II. In WoS there is only one journal satisfying the combined criteria, whereas in Scopus there are 32.

Our results suggest that WoS and especially Scopus tend to be too lenient in assigning journals to categories. A significant share of the journals in both databases, but especially in Scopus, seem to have assignments to too many categories. The databases could adopt a stricter policy in assigning journals to categories. Such a policy could be supported by the use of citation analysis.

In addition to our main findings summarized above, there are two points worth emphasizing. First, Scopus sometimes has confusing category labels. In particular, Scopus sometimes has two categories with very similar labels. Examples are the categories LINGUISTICS & LANGUAGE and LANGUAGE & LINGUISTICS and the categories INFORMATION SYSTEMS & MANAGEMENT and MANAGEMENT INFORMATION SYSTEMS. This problem could be addressed either by merging categories with similar labels or by improving the labels of these categories to make sure the differences between the categories are more clear. Second, lack of transparency is a weakness of both the WoS and the Scopus classification system. We did not find proper documentation of the methods used to construct and update the WoS and Scopus classification systems.

*6.2. Limitations and future research*

It should be emphasized that our analysis is based only on direct citation relations between journals and categories. As already mentioned, other non-citation-based approaches, in particular text-based and expert-based approaches, could also be used for assessing the accuracy of journal classification systems. These approaches are probably more effective for journals with only a small number of citation relations, for instance newly established journals. In this paper, we did not take non-citation-based approaches into consideration. Hence, when we conclude that the assignment of a journal is questionable, one should be aware that this conclusion is drawn purely from a citation perspective. In some cases, another perspective may lead to a different conclusion. For instance, our citation perspective suggests that *Australian Journal of Management*, discussed in Subsection 5.3, is misclassified in WoS, but an expert judgment based on the scope statement of the journal may result in a different conclusion.

Furthermore, when a citation-based approach is taken, the effectiveness of the use of direct citation relations might be questioned in some fields of science. This is the case especially in fields in which scientific journals play a less significant role and in which sources such as books, which are not properly covered in WoS and Scopus, are more important. By considering only direct citation relations between journals, a significant share of the scientific communication in these fields is ignored, which might have a negative effect on the accuracy of our analysis. Other citation-based approaches, for instance using bibliographic coupling relations instead of direct citation relations, may offer a solution. Two journals belonging to the same category may for instance have hardly any direct citation relations with each other, but they may refer a lot to the same books, and therefore they may have many bibliographic coupling relations. To further explore this possibility, we tested a bibliographic coupling approach in the WoS category CULTURAL STUDIES. Using a direct citation approach, 61% of the journals in the CULTURAL STUDIES category satisfy Criterion I ($\alpha = 0.1$; see Table 4). Using a bibliographic coupling approach, it turns out that a similar result is obtained.[10] Hence, we have no clear evidence to support the idea that in some fields a bibliographic coupling approach may be more suitable than a direct citation approach.

In the case of a direct citation approach, it could be suggested that in the calculation of the relatedness between a journal and a category only the outgoing citations of a journal should be considered instead of both the incoming and the outgoing citations. Certain journals, for instance journals focused on methodological topics, may be cited by journals from many different categories. For these journals, it might perhaps be better to consider only their outgoing citations in the calculation of relatedness. This may be worth studying in future research.

Additionally, as already mentioned in Section 2, WoS and Scopus differ in their coverage of scientific literature. It could be argued that differences in coverage may have an effect on our analysis. For instance, if Scopus covers relatively more journals than WoS in research fields in which it is relatively difficult to produce an accurate journal classification, then this could to some degree explain why in our analysis the classification system of Scopus appears to be less accurate than the WoS classification system. On the other hand, Mongeon and Paul-Hus (2016) indicate that, even though Scopus has a broader coverage than WoS in all fields of science, the two databases have similar biases in their coverage of fields. In our analysis, we have focused mainly on relative rather than absolute statistics (e.g., the percentage of journals satisfying a criterion rather than the absolute number of journals satisfying a criterion). In this way, we have corrected for the fact that Scopus covers more journals than WoS. However, we did not correct for possible differences between WoS and Scopus in the distribution of journals over fields. The effect of such differences could be examined in future research by comparing the classification accuracy of WoS and Scopus at the level of individual fields.

Another topic for future research could be the issue of differences in the size of categories. Some categories are much larger than others in terms of their number of journals and publications. This has certain consequences for our analysis. For instance, in the case of a small category, it may be hardly possible for a journal to have a reasonably high relatedness with the category. Therefore it can be expected that many journals belonging to the category will satisfy Criterion I. This may be caused not so much by the misclassification of these journals but more by the small size of the category. On the other hand, in the case of a large category, there may be other problems. A large category may for instance be of a heterogeneous nature and may cover multiple fields that are hardly connected to each other. Our Criteria I and II are unable to detect this problem. The issue of category size may be studied in more detail in future research.

Finally, it should be noted that the usefulness of journal classification systems can be questioned at a more fundamental level. Multidisciplinary journals such as *Nature*, *PNAS*, and *Science* do not fit well in a classification system at the level of journals. With the increasing popularity of large multidisciplinary open access journals, such as *PLoS ONE* and *Scientific Reports*, this problem is becoming more and more serious. An increasing share of all scientific publications appear in multidisciplinary journals, and these publications cannot be properly classified at the journal level. This makes it increasingly important to develop multidisciplinary classification systems at the level of individual publications rather than at the journal level. Algorithmic approaches to construct such publication-level classification systems have been studied in a number of recent papers (Boyack & Klavans, 2010; Boyack et al., 2011; Klavans & Boyack, 2015; Waltman & Van Eck, 2012). The resulting classification systems are likely to play an important role in future bibliometric analyses.

---

[10] 38 journals are assigned to the CULTURAL STUDIES category in WoS, of which there are 28 with $t_i \geq 100$. For each of these 28 journals, we selected the category with which the journal has most bibliographic coupling relations or most direct citation relations. Based on both bibliographic coupling relations and direct citation relations, it is found that none of the 28 journals has CULTURAL STUDIES as the category with which it is most strongly connected; instead, the 28 journals are more strongly connected to categories such as SOCIOLOGY, GEOGRAPHY, COMMUNICATION, and ANTHROPOLOGY. Based on this finding, we conclude that a bibliographic coupling approach and a direct citation approach yield similar results in the case of the CULTURAL STUDIES category.

## Acknowledgements

We would like to thank Ulf Sandström and Ismael Rafols for their comments on earlier drafts of this paper and Nees Jan van Eck for his helpful suggestions regarding the use of the Web of Science and Scopus databases. We are grateful to participants in the CWTS research seminar for their feedback on this research. We also thank two reviewers for their comments on this paper.

## Appendix

Table A1. WoS multidisciplinary categories

| |
|---|
| AGRICULTURE, MULTIDISCIPLINARY |
| CHEMISTRY, MULTIDISCIPLINARY |
| COMPUTER SCIENCE, INTERDISCIPLINARY APPLICATIONS |
| ENGINEERING, MULTIDISCIPLINARY |
| GEOSCIENCES, MULTIDISCIPLINARY |
| HUMANITIES, MULTIDISCIPLINARY |
| MATERIALS SCIENCE, MULTIDISCIPLINARY |
| MATHEMATICS, INTERDISCIPLINARY APPLICATIONS |
| MEDICINE, GENERAL & INTERNAL |
| MULTIDISCIPLINARY SCIENCES |
| PHYSICS, MULTIDISCIPLINARY |
| PSYCHOLOGY, MULTIDISCIPLINARY |
| SOCIAL SCIENCES, INTERDISCIPLINARY |

Table A2. Journals satisfying both Criterion I and Criterion II (Scopus; $\alpha = 0.1$; $\beta = 0.6$; excluding journals with $t_i < 100$)

| Scopus journal | Criterion I | | Criterion II | |
|---|---|---|---|---|
| | Scopus category | $r_{i,c}$ | Scopus category | $r_{i,c}$ |
| *Analog Integrated Circuits and Signal Processing* | HARDWARE AND ARCHITECTURE | 0.06 | ELECTRICAL AND ELECTRONIC ENGINEERING | 0.82 |
| *Analog Integrated Circuits and Signal Processing* | SIGNAL PROCESSING | 0.05 | ELECTRICAL AND ELECTRONIC ENGINEERING | 0.82 |
| *Analog Integrated Circuits and Signal Processing* | SURFACES, COATINGS AND FILMS | 0.05 | ELECTRICAL AND ELECTRONIC ENGINEERING | 0.82 |
| *Ancient Mesoamerica* | GEOGRAPHY, PLANNING AND DEVELOPMENT | 0.01 | ARCHEOLOGY | 0.69 |
| *Asian Perspective* | LIFE-SPAN AND LIFE-COURSE STUDIES | 0.00 | POLITICAL SCIENCE AND INTERNATIONAL RELATIONS | 0.61 |
| *Caikuang yu Anquan Gongcheng Xuebao/Journal of Mining and Safety Engineering* | SAFETY, RISK, RELIABILITY AND QUALITY | 0.02 | GEOTECHNICAL ENGINEERING AND ENGINEERING GEOLOGY | 0.80 |
| *Clinical Research in Cardiology* | MOLECULAR BIOLOGY | 0.05 | CARDIOLOGY AND | 0.62 |
| *Supplements* | | | CARDIOVASCULAR MEDICINE | |
| *Clinical Research in Cardiology Supplements* | RADIOLOGY, NUCLEAR MEDICINE AND IMAGING | 0.05 | CARDIOLOGY AND CARDIOVASCULAR MEDICINE | 0.62 |
| *Clinical Research in Cardiology Supplements* | STRUCTURAL BIOLOGY | 0.00 | CARDIOLOGY AND CARDIOVASCULAR MEDICINE | 0.62 |
| *Computer Graphics Forum* | COMPUTER NETWORKS AND COMMUNICATIONS | 0.01 | COMPUTER GRAPHICS AND COMPUTER-AIDED DESIGN | 0.64 |
| *Cooperation and Conflict* | STRATEGY AND MANAGEMENT | 0.01 | POLITICAL SCIENCE AND INTERNATIONAL RELATIONS | 0.67 |
| *Cultural Studies of Science Education* | CULTURAL STUDIES | 0.04 | EDUCATION | 0.77 |
| *Current Bladder Dysfunction Reports* | BIOCHEMISTRY | 0.00 | UROLOGY | 0.70 |
| *Current Bladder Dysfunction Reports* | MOLECULAR BIOLOGY | 0.01 | UROLOGY | 0.70 |
| *Current Cardiovascular Imaging Reports* | APPLIED MICROBIOLOGY AND BIOTECHNOLOGY | 0.00 | CARDIOLOGY AND CARDIOVASCULAR MEDICINE | 0.65 |
| *Current Cardiovascular Imaging Reports* | CELL BIOLOGY | 0.00 | CARDIOLOGY AND CARDIOVASCULAR MEDICINE | 0.65 |
| *Current Cardiovascular Imaging Reports* | HISTOLOGY | 0.00 | CARDIOLOGY AND CARDIOVASCULAR MEDICINE | 0.65 |
| *Economics of Governance* | BUSINESS AND INTERNATIONAL MANAGEMENT | 0.03 | ECONOMICS AND ECONOMETRICS | 0.72 |
| *Federal Reserve Bank of St. Louis Review* | BUSINESS AND INTERNATIONAL MANAGEMENT | 0.02 | ECONOMICS AND ECONOMETRICS | 0.68 |
| *Filozofia* | RELIGIOUS STUDIES | 0.06 | PHILOSOPHY | 0.63 |
| *Geotechnique Letters* | ATMOSPHERIC SCIENCE | 0.03 | GEOTECHNICAL ENGINEERING AND ENGINEERING GEOLOGY | 0.60 |
| *Handbook of Social Economics* | SOCIOLOGY AND POLITICAL SCIENCE | 0.08 | ECONOMICS AND ECONOMETRICS | 0.66 |
| *Higher Education* | LAW | 0.03 | EDUCATION | 0.63 |
| *International Journal of Dynamical Systems and Differential Equations* | CONTROL AND OPTIMIZATION | 0.02 | APPLIED MATHEMATICS | 0.62 |
| *International Journal of Dynamical Systems and Differential Equations* | DISCRETE MATHEMATICS AND COMBINATORICS | 0.05 | APPLIED MATHEMATICS | 0.62 |
| *International Journal of Geomechanics* | SOIL SCIENCE | 0.09 | GEOTECHNICAL ENGINEERING AND ENGINEERING GEOLOGY | 0.65 |
| *Journal of Cryptology* | APPLIED MATHEMATICS | 0.09 | THEORETICAL COMPUTER SCIENCE | 0.64 |
| *Journal of Cryptology* | COMPUTER SCIENCE APPLICATIONS | 0.08 | THEORETICAL COMPUTER SCIENCE | 0.64 |
| *Journal of Cryptology* | SOFTWARE | 0.09 | THEORETICAL COMPUTER SCIENCE | 0.64 |
| *Journal of Personal Selling and Sales Management* | HUMAN FACTORS AND ERGONOMICS | 0.00 | MARKETING | 0.66 |
| *Journal of Personal Selling and Sales Management* | MANAGEMENT OF TECHNOLOGY AND INNOVATION | 0.08 | MARKETING | 0.66 |
| *Managerial Auditing Journal* | ORGANIZATIONAL BEHAVIOR AND HUMAN RESOURCE MANAGEMENT | 0.06 | ACCOUNTING | 0.62 |
| *Memoirs of the Queensland Museum* | ECOLOGY | 0.08 | ECOLOGY, EVOLUTION, BEHAVIOR AND SYSTEMATICS | 0.61 |
| *Memoirs of the Queensland Museum* | PALEONTOLOGY | 0.03 | ECOLOGY, EVOLUTION, BEHAVIOR AND SYSTEMATICS | 0.61 |
| *Mobilization* | TRANSPORTATION | 0.00 | SOCIOLOGY AND POLITICAL SCIENCE | 0.64 |
| *Multicultural Perspectives* | CULTURAL STUDIES | 0.09 | EDUCATION | 0.77 |
| *Perspektiven der Wirtschaftspolitik* | GEOGRAPHY, PLANNING AND DEVELOPMENT | 0.08 | ECONOMICS AND ECONOMETRICS | 0.60 |
| *Perspektiven der Wirtschaftspolitik* | POLITICAL SCIENCE AND INTERNATIONAL RELATIONS | 0.03 | ECONOMICS AND ECONOMETRICS | 0.60 |
| *Portal: Libraries and the Academy* | DEVELOPMENT | 0.00 | LIBRARY AND INFORMATION SCIENCES | 0.83 |
| *Public Services Quarterly* | ACCOUNTING | 0.00 | LIBRARY AND INFORMATION | 0.80 |

| | | | SCIENCES | |
|---|---|---|---|---|
| *Public Services Quarterly* | PUBLIC ADMINISTRATION | 0.03 | LIBRARY AND INFORMATION SCIENCES | 0.80 |
| *RAIRO - Theoretical Informatics and Applications* | COMPUTER SCIENCE APPLICATIONS | 0.07 | THEORETICAL COMPUTER SCIENCE | 0.69 |
| *RAIRO - Theoretical Informatics and Applications* | SOFTWARE | 0.01 | THEORETICAL COMPUTER SCIENCE | 0.69 |
| *Review of Political Economy* | POLITICAL SCIENCE AND INTERNATIONAL RELATIONS | 0.07 | ECONOMICS AND ECONOMETRICS | 0.64 |
| *Revue d'Economie Politique* | POLITICAL SCIENCE AND INTERNATIONAL RELATIONS | 0.04 | ECONOMICS AND ECONOMETRICS | 0.60 |
| *State Politics and Policy Quarterly* | ARTS AND HUMANITIES (MISCELLANEOUS) | 0.01 | SOCIOLOGY AND POLITICAL SCIENCE | 0.69 |
| *State Politics and Policy Quarterly* | POLITICAL SCIENCE AND INTERNATIONAL RELATIONS | 0.06 | SOCIOLOGY AND POLITICAL SCIENCE | 0.69 |